# Secure Degrees of Freedom for Gaussian Channels with Interference: Structured Codes Outperform Gaussian Signaling

Xiang He    Aylin Yener
Wireless Communications and Networking Laboratory
Electrical Engineering Department
The Pennsylvania State University, University Park, PA 16802
*xxh119@psu.edu*    *yener@ee.psu.edu*

***Abstract*—In this work, we prove that a positive secure degree of freedom is achievable for a large class of Gaussian channels as long as the channel is not degraded and the channel is fully connected. This class includes the MAC wire-tap channel, the 2-user interference channel with confidential messages, the 2-user interference channel with an external eavesdropper. Best known achievable schemes to date for these channels use Gaussian signaling. In this work, we show that structured codes outperform Gaussian random codes at high SNR when channel gains are real numbers.**

## I. Introduction

Information theoretic security, originally proposed by Shannon [1], seeks the fundamental limits of reliable transmission rates when the messages must be kept secret from a computation-power unlimited adversary whose observation of the transmitted signals contain some uncertainty. By now, it is well known that introducing interference into the channel in a proper manner may increase the uncertainty observed by the adversary and hence allow for a higher rate of secret messages [2]–[4]. The interference should be introduced in a way such that it is more harmful to the adversary than it is to the intended receiver of the messages. Hence, the key is to achieve a fine balance between secrecy against the adversary and the level of harmful interference to the system.

For these channel models, the achieved rate obtained so far is far from the outer bounds. For example, the genie outer bound from [4] increases with power $P$ at the speed of $0.5\log_2(P)$ [5, (69)]. The achievable secrecy rate converges to a constant when $P$ goes to $\infty$ [4, Theorem 2]. This means the gap between the achievable rate and the outer bound is unbounded and the trade-off between secrecy and interference is still not well-understood. In fact, once the channel model is such that the intended receiver is not harmed by the introduced interference, the achieved secrecy rate immediately comes within $0.5$ bits/channel use of the capacity region, as was shown for the one sided interference channel in [6] and the orthogonal MAC wire-tap channel in [7].

In this work, we consider the more general case where introducing interference will both confuse the eavesdropper and harm the intended receiver simultaneously. We show, for a large class of Gaussian channels with confidential messages where introducing interference effect both the intended receiver and the eavesdropper, that signaling using structured codes can out-perform signaling with i.i.d. Gaussian codebooks in high SNR. This class includes the Gaussian MAC-wiretap channel [2], the Gaussian interference channel with confidential messages [3] and the Gaussian interference channel with an external eavesdropper [7]. It has been a folk conjecture that the achievable rate regions with Gaussian codebooks in these works were likely optimal and efforts [4], [5], [7] have been made to find outer bounds to prove this. A direct consequence of the result we report in this paper is that this is not so.

This insight comes from studying the secure degree of freedom of the interference assisted wire-tap channel [4], which falls under the three channels mentioned above when only one source node has confidential message to send. As mentioned before, reference [4] shows that the achieved rate using Gaussian codebooks converges to a constant as power increase, which implies the obtained secure degree of freedom for this channel is $0$. In contrast, we find that a positive degree of freedom is actually achievable for all channel gains as long as the channel is not degraded. The key to getting this result is the use of different types of structured codes for appropriate channel gains rather than Gaussian signaling.

We note that a positive secure degree of freedom is known to be achievable for the fading channel [8], which requires coding over different fading states and hence does not imply the result here.

The result here provides another example that structured codes are useful in proving information theoretic results. A list of examples that structured codes outperform simple random coding arguments in non-secrecy problems can be found in [9]. Using structured code in secrecy problems was first proposed by the authors in [10]. Up to date structured codes are found to be useful for relay channels due to the possibility of compute-and-forward [9], [10], or for interference channels with more than two users due to the possibility of interference alignment [11]–[13]. The result here provides the first example that structured codes are indeed useful for two user Gaussian

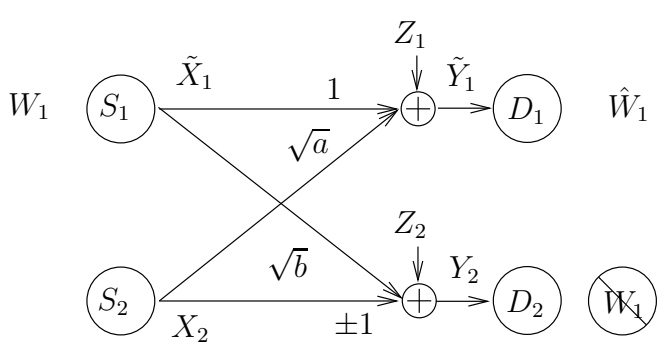

Fig. 1. Interference-assisted Wire-tap Channel

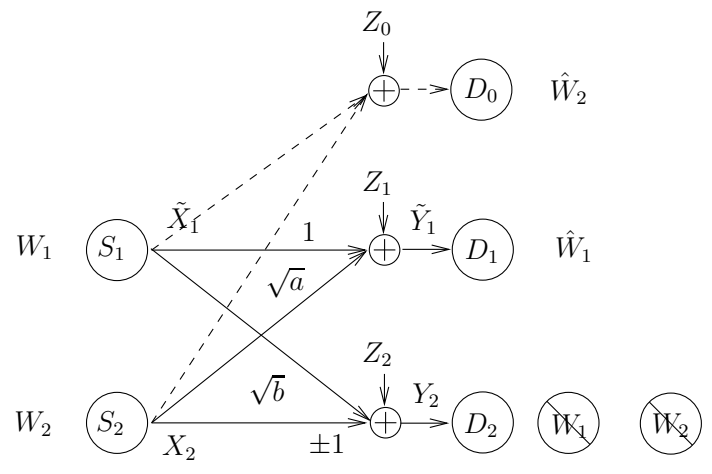

Fig. 2. Interference-assisted Wire-tap Channel as a Special Case of the Interference Channel with an External Eavesdropper

channels as well.

## II. System Model

Consider the Gaussian interference-assisted wire-tap channel [4] shown in Figure 1. In this model, node $S_1$ sends a secret message $W_1$ via $\tilde{X}_1$ to node $D_1$, which must be kept secret from node $D_2$. We assume the channel is fully connected, which means that no link's channel gain equals zero. This assumption is obviously valid for a wireless medium. Then after normalizing the channel gains of the two intended links to 1, the received signals at the two receiving node $D_1$ and $D_2$ can be expressed as

$$\begin{aligned} \tilde{Y}_1 &= \tilde{X}_1 + \sqrt{a} X_2 + Z_1 \\ Y_2 &= \sqrt{b}\tilde{X}_1 \pm X_2 + Z_2 \end{aligned} \tag{1}$$

where $Z_i, i = 1,2$ is a zero-mean Gaussian random variable with unit variance. For now, we assume $\sqrt{a}$ and $\sqrt{b}$ are real numbers. The case with complex numbers will be briefly explained in Section II-B.

Since $W_1$ must be kept secret from $D_2$, we require

$$\lim_{n\to\infty} \frac{1}{n} H\left(W_1\right) = \lim_{n\to\infty} \frac{1}{n} H\left(W_1 | Y_2^n\right) \tag{2}$$

The achieved secrecy rate $R_e$ is defined as $\lim_{n\to\infty} \frac{1}{n} H\left(W_1\right)$ such that the condition (2) is fulfilled and $W_1$ can be reliably received by $D_1$.

Let $X_1 = \sqrt{b}\tilde{X}_1$ and $Y_1 = \sqrt{b}\tilde{Y}_1$. Then from (1), we have

$$\begin{aligned} Y_1 &= X_1 + \sqrt{ab} X_2 + \sqrt{b} Z_1 \\ Y_2 &= X_1 \pm X_2 + Z_2 \end{aligned} \tag{3}$$

In the sequel, we will focus on this scaled model instead, as we find it more convenient to use it to explain our results.

Let the average power constraint of node $S_i$ on $X_i$ be $\bar{P}_i$. The secure degree of freedom of the secrecy rate is defined as

$$\limsup_{\bar{P}_i\to\infty, i=1,2} \frac{R_e}{\frac{1}{2}\log_2\left(\sum_{i=1}^{2} \bar{P}_i\right)} \tag{4}$$

It is clear that the secure degree does not change, whether the model is described via (1) or (3).

### A. Relationship with Other Channels

The significance of the interference-assisted wiretap channel is that it can be considered as a special case of a large class of channel models with confidential messages, as shown below:

1) If node $S_2$ has a confidential message $W_2$ for $D_1$, which must be kept secret from $D_2$, then the channel is the MAC-wiretap channel considered in [2].
2) If node $S_2$ has a confidential message $W_2$ for $D_2$, and the message must be kept secret from $D_1$, then the channel is the interference channel with confidential message considered in [3].
3) As shown in Figure 2, we can add another receiving node $D_0$ to Figure 1, to which node $S_2$ wants to sent a confidential message $W_2$. Again $W_2$ must be kept secret from $D_2$. Then the channel becomes the interference channel with an external eavesdropper considered in [7, Section VI].

Hence, any secrecy rate achieved in the the interference-assisted wire-tap channel is an achievable *individual rate* for all the three multi-user channels mentioned above.

*Remark 1:* The results here also strengthen a result the authors derived previously for the $K$-user interference channel with confidential messages, where $K \geq 3$ [11]. In [11], it is not known if, for sum secrecy rate, a positive secure degree of freedom is achievable for *arbitrary* channel gains. Since the interference-assisted wire-tap channel is also a special case of the $K$-user channel, we see the answer to this question is yes unless the channel is degraded for any pair of the $K$ users.

### B. Complex Channel Gains

More general than the channel with real channel gains is the channel with complex channel gains. The reason that we focus on the real case in the sequel is that the complex case is actually easier in terms of achieving positive secure degree of freedom, as explained below.

Since the channel is fully connected, after normalization of the channel gains and variable substitution, the received signals at nodes $D_1$ and $D_2$ can be expressed as [14]:

$$\begin{aligned} Y_1 &= X_1 + \sqrt{ab} e^{j\psi} X_2 + \sqrt{b} Z_1 \\ Y_2 &= X_1 + X_2 + Z_2 \end{aligned} \tag{5}$$

where $Z_i, i = 1,2$ are rotational invariant complex Gaussian random variables with unit variance. Then we have:

*Theorem 1:* A secure degree of freedom of $1$ is achievable if $\psi \neq 0$ or $\pi \mod 2\pi$.

*Proof Outline:* Let $\mathrm{Im}X_i = 0, i = 1,2$. Let $\cot x = \cos x / \sin x$. Then since $\mathrm{Im}Y_2 = \mathrm{Im}Z_2$, $\mathrm{Im}Y_2$ does not provide any information about $W_1$ to the eavesdropper. Hence we can assume the eavesdropper receives $\mathrm{Re}Y_2$ only. Node $D_1$ computes $g(Y_1) = \mathrm{Re}Y_1 - \cot\psi \mathrm{Im}Y_1$. Then the channel can

be expressed as

$$g(Y_1) = \text{Re}X_1 + \sqrt{b}\left(\text{Re}Z_1 - \cot\psi \text{Im}Z_1\right)$$
$$\text{Re}Y_2 = \text{Re}X_1 + \text{Re}X_2 + \text{Re}Z_2 \tag{6}$$

The channel then becomes an one-sided interference channel. By transmitting a i.i.d Gaussian noise via $\text{Re}X_2$, the channel is equivalent to a Gaussian wire-tap channel. It is known that the following secrecy rate is achievable [15]:

$$C\left(\frac{P_1}{(b\csc^2\psi)/2}\right) - C\left(\frac{P_1}{P_2 + 1/2}\right) \tag{7}$$

where $C(x) = \frac{1}{2}\log_2(1 + x)$. $P_i$ is the average power constraint on $X_i$. Hence a secure degree of freedom of 1 is achievable for this channel. ■

### C. Gaussian Signaling

In [4], an achievable rate is derived with Gaussian codebooks and power control. One implication of this achievable rate is the high SNR behavior as described by Theorem 2 therein. We re-state this result below:

*Theorem 2:* With Gaussian codebooks, the achievable secrecy rate $R_1$ converges to a constant when the power constraint of node $D_1$ and $D_2$ goes to $\infty$.

This means the achieved secure degree of freedom by the coding scheme in [4] is 0.

## III. The Achievable Scheme

### A. Results on Structured Codes

*1) Nested Lattice:* A nested lattice code is defined as an intersection of $N$-dimensional fine lattice $\Lambda$ and the fundamental region of an $N$-dimensional "coarse" lattice $\Lambda_c$, denoted by $\mathcal{V}(\Lambda_c)$. We require that $\Lambda_c \subset \Lambda$. Let $u_i^N$ be uniformly distributed over $\Lambda \cap \mathcal{V}(\Lambda_c)$. Let the dithering noise $d_i^N$ be a continuous random vector which is uniformly distributed over $\mathcal{V}(\Lambda_c)$. Define modulus operation such that $x \bmod \Lambda_c = x - arg\min_{u\in\Lambda_c} \|x - u\|$. Then the values of $X_i$ over $N$ channel uses are computed as

$$X_i^N = (u_i^N + d_i^N) \bmod \Lambda_c \tag{8}$$

$u_i^N, d_i^N, i = 1, 2$ are independent. We also assume $d_i^N, i = 1, 2$ are known by all receiving nodes. Hence they are not used to enhance secrecy.

As will be shown later, we are interested in lower-bounding the expression $I(u_1^N; X_1^N \pm X_2^N, d_1^N, d_2^N)$, which corresponds to the rate of information leaked to the eavesdropper. To do that, we need the following result. Its proof follows from the representation theorem introduced in [10] and is given in [14].

*Theorem 3:* There exists an random integer $T$, such that $1 \le T \le 2^N$, and $X_1^N \pm X_2^N$ is uniquely determined by $\{T, X_1^N \pm X_2^N \bmod \Lambda_c\}$.

Using Theorem 3, we have

$$I(u_1^N; X_1^N \pm X_2^N, d_1^N, d_2^N)$$
$$= I(u_1^N; X_1^N \pm X_2^N \bmod \Lambda_c, T, d_1^N, d_2^N) \tag{9}$$
$$\le I(u_1^N; X_1^N \pm X_2^N \bmod \Lambda_c, d_1^N, d_2^N) + H(T) \tag{10}$$
$$= I(u_1^N; u_1^N \pm u_2^N \bmod \Lambda_c) + H(T) \tag{11}$$
$$= H(T) \le N \tag{12}$$

This means at most $N$ bit per channel use is leaked to the eavesdropper over $N$ channel uses.

*2) Integer Lattice:* An integer lattice code with parameter $Q$ is composed of points in the set $[0, Q) \cap \mathbf{Z}$ where $\mathbf{Z}$ is the set of all integers. As will be shown later, in this case, the rate of information leaked to the eavesdropper is given by $f(Q)$ defined as:

$$f(Q) = I(X_1; X_1 \pm X_2) \tag{13}$$

where $X_i, i = 1, 2$ is uniformly distributed over $[0, Q) \cap \mathbf{Z}$. $f(Q)$ can be lower bounded by the following lemma:

*Lemma 1:* For a positive integer $Q$,

$$f(Q) \le \frac{1}{2}\log_2(2\pi e(\frac{1}{6} - \frac{1}{12Q^2})) < \frac{1}{2}\log_2(\frac{\pi e}{3}) < 0.8 \tag{14}$$

The proof follows from [13, Lemma 12] and is given in [14].

We next use these results to derive achievable secure degree of freedom for the interference assisted wire-tap channel.

### B. When $\sqrt{ab}$ is algebraic irrational

*Theorem 4:* A secure degree of freedom of $1/2$ is achievable when $\sqrt{ab}$ is an algebraic irrational number.

*Proof:* We use the lattice codebook used in [13, Theorem 1]. Let $\Lambda_{P,\varepsilon}$ be the scalar lattice defined as:

$$\Lambda_{P,\varepsilon} = \left\{x : x = P^{1/4+\varepsilon}z, z \in \mathbf{Z}\right\} \tag{15}$$

The codebook $\mathcal{C}_{P,\varepsilon}$ is given by:

$$\mathcal{C}_{P,\varepsilon} = \Lambda_{P,\varepsilon} \cap \left[-\sqrt{P}, \sqrt{P}\right] \tag{16}$$

where $P = \min\{\bar{P}_1, \bar{P}_2\}$. It then can be verified that, for large enough $P$, we have

$$\log_2|\mathcal{C}_{P,\varepsilon}| \ge \log_2\left(2P^{1/4-\varepsilon} - 1\right) \ge \log_2\left(P^{1/4-\varepsilon}\right) \tag{17}$$

The codebook is used for both node $S_1$ and node $S_2$. The codeword transmitted by node $S_1$ is chosen based on the secret message $W_1$. The codeword transmitted by node $S_2$ is chosen independently according to a uniform distribution.

Since the input from $S_2$ is i.i.d., the channel is then equivalent to a memoryless wire-tap channel [16]. According to [16], any secrecy rate $R$ such that

$$R < I(X_1; Y_1) - I(X_1; Y_2) \tag{18}$$

is achievable. Hence we need to find a lower bound to the right hand side of (18).

According to [13, Theorem 1], $p(X_1)$ is chosen to be a uniform distribution over $\mathcal{C}_{P,\varepsilon}$. Under this input distribution, following a similar derivation to [13, Theorem 1], it can be shown that when $P > \frac{1}{\alpha^2\beta^2}$, we have

$$I(X_1; Y_1) \ge \left(1 - 2\exp\left(-\frac{P^{2\varepsilon}}{8b}\right)\right)\log_2(|\mathcal{C}_{P,\varepsilon}|) - 1 \tag{19}$$

For $I(X_1; Y_2)$, we have

$$I(X_1; Y_2) \leq I(X_1; Y_2, Z_2) = I(X_1; X_1 \pm X_2) \leq 0.8 \quad (20)$$

where (20) follows from Lemma 1. Using (19) (20), and (17), we find (18) is lower bounded by

$$\left(1 - 2\exp\left(-\frac{P^{2\varepsilon}}{8b}\right)\right)\left(\frac{1}{4} - \varepsilon\right)\log_2(P) - 1.8 \quad (21)$$

for sufficiently large $P$. From (17), $\varepsilon$ can take any value between $(0, 1/4)$. Hence we have completed the proof. ■

*Remark 2:* When $\sqrt{ab} = 1$ and all channel gains are positive, the channel is degraded and from the outer bound in [4], the secure degree of freedom is 0. Since algebraic irrational numbers are dense on the real line, it follows that the secure degree of freedom is discontinuous at $\sqrt{ab} = 1$.

The result in Section III-B only applies when $\sqrt{ab}$ is algebraic irrational, which is a set of measure 0 on the real line. In the sequel we consider the case where $\sqrt{ab}$ is either rational or transcendental.

### C. When $\sqrt{ab} \geq 2$ or $1/\sqrt{ab} \geq 1/2$

Here we use the $Q$-bit expansion scheme similar to the one in [17]. Let $Q = \sqrt{ab}$ if $\sqrt{ab} \geq 2$. Otherwise, let $Q = 1/\sqrt{ab}$. Let $\lfloor Q \rfloor$ denote the largest integer $\leq Q$.

*Theorem 5:* The following secure degree of freedom is achievable:

$$\frac{1}{2}\frac{\log_2 \lfloor Q \rfloor}{\log_2 Q} - \frac{f(\lfloor Q \rfloor)}{2\log_2 Q} \quad (22)$$

where $f(Q)$ is defined in (13). (22) is lower bounded by

$$\frac{1}{2}\frac{\log_2 \lfloor Q \rfloor}{\log_2 Q} - \frac{\log_2\left(2\pi e\left(\frac{1}{6}\right) - \frac{1}{12\lfloor Q \rfloor^2}\right)}{4\log_2(Q)} \quad (23)$$

For $Q = 2$, (22) equals 0.25.

*Proof:* We begin by considering the case when $\sqrt{ab} \geq 2$.

$$X_k = \sqrt{P_0}\sum_{i=0}^{M-1} a_{k,i} Q^{2i}, k = 1, 2 \quad (24)$$

where $P_0$ is a constant scaling factor. $a_{k,i}$ is uniformly distributed over $[0, \lfloor Q \rfloor - 1] \cap \mathbf{Z}$, hence $a_{k,i}$ is uniquely determined by $X_k$.

The signal received by node $D_1$ is given by

$$Y_1 = \sqrt{P_0}\left(\sum_{i=0}^{M-1} a_{1,i} Q^{2i} + \sum_{i=0}^{M-1} a_{2,i} Q^{2i+1}\right) + \sqrt{b} Z_1 \quad (25)$$

We then derive a lower bound to $I(X_1; Y_1) - I(X_2; Y_2)$ as we did for Theorem 4.

Following a similar derivation to [13, Theorem 1], with Fano's inequality, it can be shown that $I(X_1; Y_1)$ is lower bounded as:

$$I(X_1; Y_1) \geq \left(1 - 2\exp\left(-\frac{P_0}{8b}\right)\right) H(X_1) - 1 \quad (26)$$

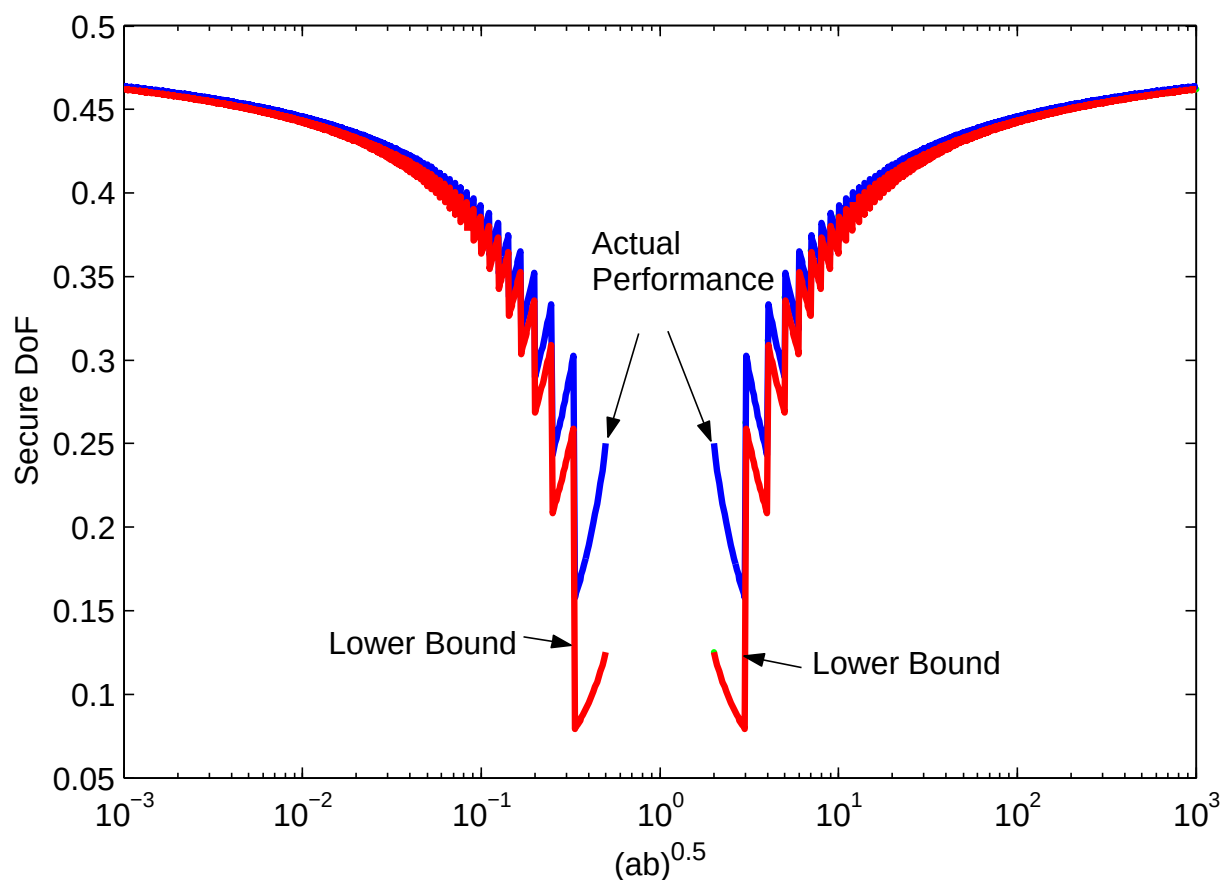

Fig. 3. Secure degree of freedom

For $I(X_1; Y_2)$, we have:

$$I(X_1; Y_2) \leq I(X_1; X_1 \pm X_2) \quad (27)$$

$$\leq \sum_{i=0}^{M-1} I(a_{1,i}; a_{1,i} \pm a_{2,i}) = M f(\lfloor Q \rfloor) \quad (28)$$

Therefore, the following secrecy rate is achievable

$$R_e = M(1 - 2\exp(-\frac{P_0}{8b}))(\log_2 \lfloor Q \rfloor) - 1 - M f(\lfloor Q \rfloor) \quad (29)$$

It can be verified that the transmission power is given by:[1]

$$Var[X_i] = P_0\left(\frac{\lfloor Q \rfloor^2 - 1}{12}\right)\frac{Q^{4M} - 1}{Q^4 - 1} \quad i = 1, 2 \quad (30)$$

The secure degree of freedom is hence given by by:

$$\lim_{M \to \infty} \frac{\left(\left(1 - 2\exp\left(-\frac{P_0}{8b}\right)\right)\log_2 \lfloor Q \rfloor - f(\lfloor Q \rfloor)\right) M}{\frac{1}{2}\log_2\left(Q^{4M}\right)} \quad (31)$$

$$= \frac{1}{2}\left(1 - 2\exp\left(-\frac{P_0}{8b}\right)\right)\frac{\log_2 \lfloor Q \rfloor}{\log_2 Q} - \frac{f(\lfloor Q \rfloor)}{2\log_2(Q)} \quad (32)$$

which can be made arbitrarily close to (22) by choosing a large enough $P_0$. (23) then follows from (22) via Lemma 1.

When $Q = 2$, it can be verified that $f(\lfloor Q \rfloor) = 0.5$, and (32) can be made to be arbitrarily close to $1/4$.

The case of $1/\sqrt{ab} \geq 2$ can be proved in a similar fashion. Details can be found in [14]. ■

In Figure 3, we plot the secure degree of freedom achieved by Theorem 5. We notice as $\sqrt{ab}$ moves away from 1, the lower bound given by (23) becomes tighter, and the secure degree of freedom converges to 0.5.

*Remark 3:* A coding scheme similar to the one described in this section can be constructed with a nested lattice code. However, the *provable* secure degree of freedom turns out to be smaller [14].

[1] In case it is desired for $X_k$ to have zero mean, we can simply shift $X_k$ by a constant, which will not change the secrecy rate.

## D. *When $\sqrt{ab} = 1$*

When $Y_2 = X_1 + X_2 + Z_2$, the channel is degraded. The secure degree of freedom is known to be 0 [4]. When $Y_2 = X_1 - X_2 + Z_2$, we have the following result:

*Theorem 6:* The secure degree of freedom of $0.1095$ is achievable.

*Proof outline:* Here we let $X_k = \sqrt{P_0}\sum_{i=0}^{M-1} a_{k,i}Q^i$. $Q = 2$. The difference is that $a_{k,i}$ is not uniformly distributed over $\{0,1\}$. Instead we choose to pick its distribution to maximize

$$I(a_{1,i}; a_{1,i} + a_{2,i}) - I(a_{1,i}; a_{1,i} - a_{2,i}) \tag{33}$$

which is about $0.1095$ when $\Pr(a_{1,i} = 1) = 0.1443$, $\Pr(a_{2,i} = 1) = 0.8557$. The theorem follows by deriving a lower bound to $I(X_1;Y_1) - I(X_2;Y_2)$. The details are provided in [14]. ■

## E. *When $1 < \sqrt{ab} < 2$ or $1/2 < \sqrt{ab} < 1$*

Let $\sqrt{ab} = p/q + \gamma/q$, where $p, q$ are coprime positive integers, and $-1 < \gamma < 1, \gamma \neq 0$. In this case, the channel can be expressed as:

$$qY_1 = qX_1 + (p+\gamma)X_2 + q\sqrt{b}Z_1 \tag{34}$$

$$Y_2 = X_1 \pm X_2 + Z_2 \tag{35}$$

*Theorem 7:* The following secure degree of freedom is achievable when $|\gamma| < 1/\sqrt{2}$:

$$\frac{\log_2\left(1-\gamma^2\right) - \log_2\left(\gamma^2\right) - 1}{\log_2\left(f(\gamma)\right) - 2\log_2\left(\gamma^2\right)} \tag{36}$$

where

$$f(\gamma) = (1-\gamma^2)(q^2 + (p+\gamma)^2) + \gamma^4 \tag{37}$$

*Proof:* Here we use a layered coding scheme similar to [11]. Let the signal send by user $k$, $X_k^N$, be the sum of codewords from $M$ layers: $X_k^N = \sum_{i=1}^{M} X_{k,i}^N$, $k = 1, 2$. For the $i$th layer, we use the nested lattice code described in Section III-A1. Let $\Lambda_i$ be the fine lattice and $\Lambda_{c,i}$ be the coarse lattice used in layer $i$. Hence the signal $X_{k,i}^N$ is computed as

$$X_{k,i}^N = \left(u_{k,i}^N + d_{k,i}^N\right) \bmod \Lambda_{c,i} \tag{38}$$

where $d_{k,i}^N$ is the dithering noise and $u_{k,i}^N$ is the lattice point:

$$u_{k,i}^N \in \mathcal{V}\left(\Lambda_{c,i}\right) \cap \Lambda_i, \quad k = 1, 2 \tag{39}$$

Define $R_i$ as the rate of the codebook for the $i$th layer. Define $P_i$ as the average power per dimension of the $i$th layer. At layer $i$, node $D_1$ first decodes $qu_{1,i}^N + pu_{2,i}^N \bmod \Lambda_{c,i}$, then decodes $u_{2,i}^N$. The decoder first computes

$$\hat{Y}_i = [\left(qu_{1,i}^N + pu_{2,i}^N\right) + \gamma X_{2,i}^N + \sum_{t=1}^{i-1}\left(qX_{1,t}^N + (p+\gamma)X_{2,t}^N\right) + q\sqrt{b}Z_1^N] \bmod \Lambda_{c,i} \tag{40}$$

Define $A_i$ as $A_i = \sum_{t=1}^{i-1}\left(q^2 + (p+\gamma)^2\right)P_t + q^2 b$. In order for node $D_1$ to correctly decode $qu_{1,i}^N + pu_{2,i}^N \bmod \Lambda_{c,i}$ from $\hat{Y}_i$, we require [14]:

$$R_i \le \frac{1}{2}\log_2\left(\frac{P_i}{\gamma^2 P_i + A_i}\right) \tag{41}$$

Node $D_1$ is then left with the following:

$$[\gamma X_{2,i}^N + \sum_{t=1}^{i-1}\left(qX_{1,t}^N + (p+\gamma)X_{2,t}^N\right) + q\sqrt{b}Z_1^N] \bmod \Lambda_{c,i} \tag{42}$$

As long as

$$P_i > \gamma^2 P_i + A_i \tag{43}$$

(42) can be approximated with high probability [14] by the following:

$$\gamma X_{2,i}^N + \sum_{t=1}^{i-1}\left(qX_{1,t}^N + (p+\gamma)X_{2,t}^N\right) + q\sqrt{b}Z_1^N \tag{44}$$

Otherwise a decoding error is said to occur. Node $D_1$ then computes:

$$[\gamma u_{2,i}^N + \sum_{t=1}^{i-1}\left(qX_{1,t}^N + (p+\gamma)X_{2,t}^N\right) + q\sqrt{b}Z_1^N] \bmod \gamma\Lambda_{c,i} \tag{45}$$

In order for node $D_1$ to decode $u_{2,i}^N$ from this signal correctly, we require [14]:

$$R_i \le \frac{1}{2}\log_2\left(\frac{\gamma^2 P_i}{A_i}\right) \tag{46}$$

Then node $D_1$ can recover the following signal from (44):

$$\sum_{t=1}^{i-1}\left(qX_{1,t}^N + (p+\gamma)X_{2,t}^N\right) + q\sqrt{b}Z_1^N \tag{47}$$

which will be fed to the decoder at lower layers.

Let the right hand side of (41) equal to the right hand side of (46):

$$\frac{P_i}{\gamma^2 P_i + A_i} = \frac{\gamma^2 P_i}{A_i} \tag{48}$$

Then, we have:

$$P_i = \alpha\left(\alpha\beta + 1\right)^{i-1} q^2 b \tag{49}$$

where $\alpha = \frac{1-\gamma^2}{\gamma^4}, \beta = q^2 + (p+\gamma)^2$. Under this power allocation, $A_i$ is given by

$$A_i = \left(\alpha\beta + 1\right)^{i-1} q^2 b \tag{50}$$

$R_i$ follows from substituting (49) and (50) into (46):

$$R_i = \frac{1}{2}\log_2\left(\frac{1-\gamma^2}{\gamma^2}\right) \tag{51}$$

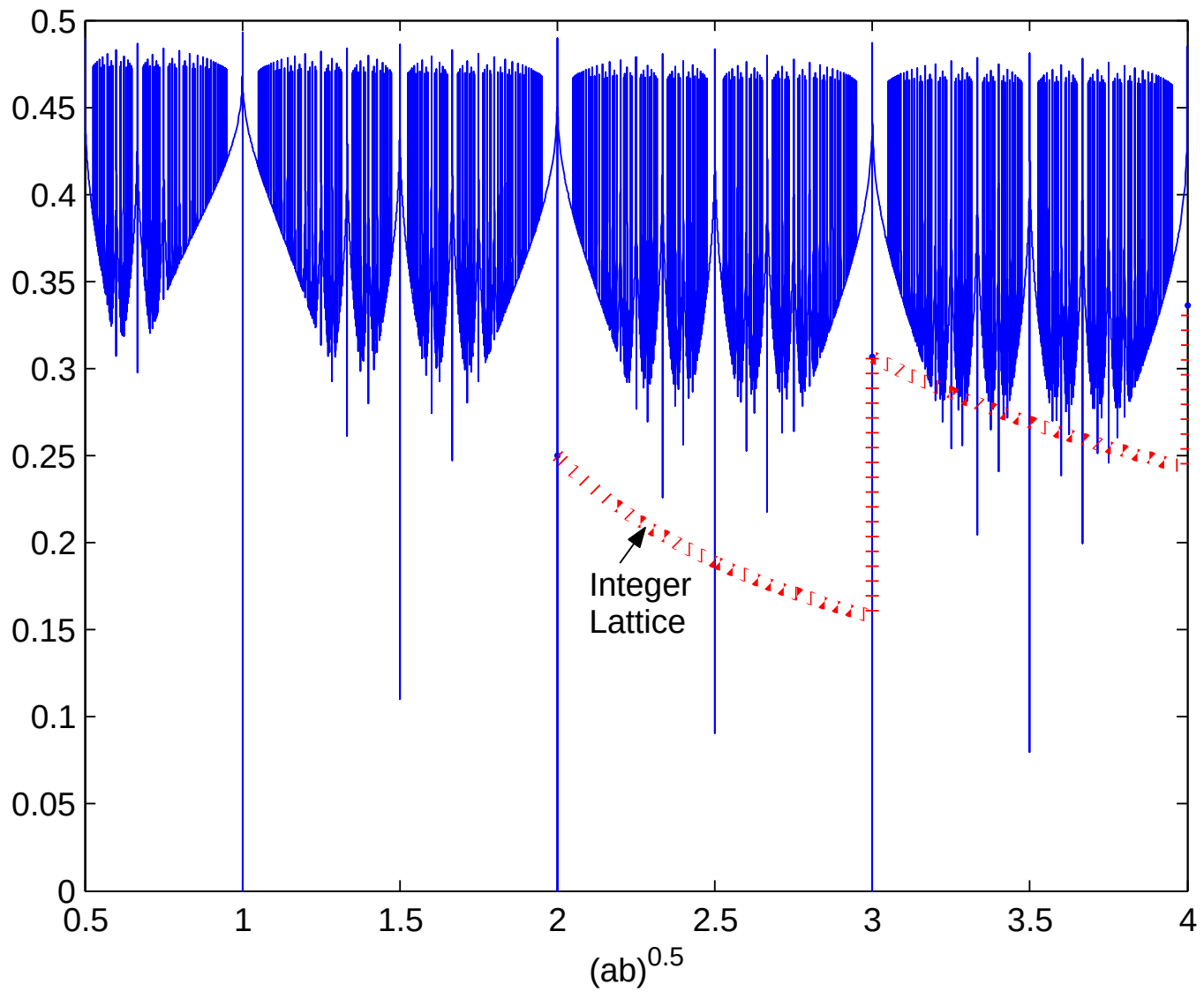

Fig. 4. Secure degree of freedom

(43) requires the term inside the log in (51) to be greater than 1. This means $|\gamma| < \frac{1}{\sqrt{2}}$. The total power of $D_i$ is given by

$$\sum_{t=1}^{M} P_t = \frac{(\alpha\beta+1)^M - 1}{\beta} q^2 b \tag{52}$$

From (12), each layer leaks at most $N$ bit to the eavesdropper over $N$ channel uses. Hence the secrecy rate $R_{e,i}$ contributed by layer $i$ is related to $R_i$ as $R_{e,i} \geq R_i - 1$. Hence a secrecy rate of $R_e = \sum_{i=1}^{M} R_i - M$ is achievable. The secure degree of freedom is therefore given by

$$\lim_{P\to\infty} \frac{R_e}{\frac{1}{2}\log_2 P} = \lim_{M\to\infty} \frac{\sum_{i=1}^{M} R_i - M}{\frac{1}{2}\log_2 P} = \frac{\frac{1}{2}\log_2\left(\frac{1-\gamma^2}{\gamma^2}\right) - 1}{\frac{1}{2}\log_2(\alpha\beta+1)} \tag{53}$$

which equals (36) in Theorem 7. ■

*Remark 4:* In Figure 4, we plot the achieved secure degree of freedom by the nested lattice coding scheme in this section. $q \leq 20$ and $p$ is chosen to be positive integers smaller than and coprime with $q$. The figure shows that the secure degree of freedom is positive when $0.5 < \sqrt{ab} < 1$ or $1 < \sqrt{ab} < 2$. This, along with the results in the previous sections, proves that *the secure degree of freedom is positive everywhere except when the channel is degraded*.

*Remark 5:* The scheme described in this section also applies to the case where $\sqrt{ab} > 2$ or $\sqrt{ab} < 1/2$. Plotted with dashed lines in Figure 4 is the performance of the integer lattice from previous section. Comparing it with the performance of the scheme in this section, we find that neither scheme dominates the other in performance.

*Remark 6:* It is possible to construct a coding scheme similar to the one described in this section using an integer lattice, which may yield a higher secure degree of freedom. However, it is difficult to find a uniform description of such code for all $p$ and $q$. Hence, instead we use a nested lattice code to prove that a positive secure degree of freedom is achievable.

## IV. Conclusion

In this work, we have proved that a positive secure degree of freedom is achievable for the fully connected interference assisted wire-tap channel when the channel is not degraded. As a consequence of this high SNR result, we are able to claim that, in contrast to common belief, Gaussian signaling is not optimal for a large class of two user Gaussian channels.

An added practical value of our result is that it implies that the cooperation of just one node is sufficient to achieve an arbitrarily large secrecy rate given enough power. Since large scale cooperation involving multiple nodes is not essential, this fact enhances the robustness of the network in an adverse environment.